\documentclass[aps,10pt, pra,notitlepage,twocolumn,superscriptaddress,longbibliography]{revtex4-2}
\usepackage{xcolor}
\usepackage{fix-cm} 
\usepackage{lmodern} 
\usepackage{times}
\usepackage{enumitem}
\usepackage{hyperref}
\hypersetup{colorlinks=true,citecolor=blue,linkcolor=blue,filecolor=blue,urlcolor=blue,breaklinks=true}
\usepackage{amsmath,amsfonts,amssymb,array,graphicx,mathtools,multirow,bm,tcolorbox,relsize,booktabs,mathrsfs}
\usepackage{newtxtext, newtxmath}
\usepackage[capitalise]{cleveref}
\usepackage{makecell}
\usepackage{pifont}
\usepackage{bbding}
\usepackage{optidef}
\usepackage{subcaption}
\usepackage{quantikz}
\usepackage{tikz}
\usepackage[normalem]{ulem} 
\usepackage{titlesec}
\usepackage{diagbox}
\titlespacing*{\section}{0pt}{\baselineskip}{\baselineskip}
\newenvironment{sketch_of_proof}{%
  \noindent{\em Sketch of proof.\ }}{%
  \hspace*{\fill}\qed
  \vspace{2ex}}

\definecolor{qblue}{rgb}{0.01,0.51,0.93}
\definecolor{qred}{rgb}{1,0,0}

\DeclarePairedDelimiter{\abs}{\lvert}{\rvert}

\usepackage{eurosym}
\usepackage[export]{adjustbox}
\usepackage{booktabs,tabularx} 
\usepackage{algorithmic}
 
\usepackage{framed}
\definecolor{shadecolor}{rgb}{0.9,0.9,0.9}

\usepackage{mathtools}
\usepackage{amsmath}

\usepackage{graphicx,epic,eepic,epsfig,amsmath,latexsym,amssymb,verbatim,color}
 
\usepackage{amsfonts}       
\usepackage{nicefrac}       

\usepackage{amsmath}
\usepackage{bbm}

\usepackage{float}
\usepackage{tikz}
\usetikzlibrary{chains}
\usetikzlibrary{fit}
\usetikzlibrary{arrows} 
\usetikzlibrary{snakes}

\usepackage{epsfig}
\usetikzlibrary{shapes.symbols,patterns} 
\usepackage{pgfplots}

\usepackage[strict]{changepage}
\usepackage{hyperref}
\hypersetup{colorlinks=true,citecolor=blue,linkcolor=blue,filecolor=blue,urlcolor=blue,breaklinks=true}

\usepackage[marginal]{footmisc}
\usepackage{url}
\usepackage{theorem}

\newtheorem{theorem}{Theorem}

\newtheorem{corollary}[theorem]{Corollary}

\def\squareforqed{\hbox{\rlap{$\sqcap$}$\sqcup$}}
\def\qed{\ifmmode\squareforqed\else{\unskip\nobreak\hfil
\penalty50\hskip1em\null\nobreak\hfil\squareforqed
\parfillskip=0pt\finalhyphendemerits=0\endgraf}\fi}
\def\endenv{\ifmmode\;\else{\unskip\nobreak\hfil
\penalty50\hskip1em\null\nobreak\hfil\;
\parfillskip=0pt\finalhyphendemerits=0\endgraf}\fi}
\newenvironment{proof}{\noindent \textbf{{Proof~} }}{\hfill $\blacksquare$}

\newcounter{remark}

\newcounter{example}

\mathchardef\ordinarycolon\mathcode`\:
\mathcode`\:=\string"8000
\def\vcentcolon{\mathrel{\mathop\ordinarycolon}}
\begingroup \catcode`\:=\active
  \lowercase{\endgroup
  \let :\vcentcolon
  }

\usepackage{cleveref}
\usepackage{graphicx}

\RequirePackage[framemethod=default]{mdframed}
\newmdenv[skipabove=7pt,
skipbelow=7pt,
backgroundcolor=darkblue!15,
innerleftmargin=5pt,
innerrightmargin=5pt,
innertopmargin=5pt,
leftmargin=0cm,
rightmargin=0cm,
innerbottommargin=5pt,
linewidth=1pt]{tBox}

\newmdenv[skipabove=7pt,
skipbelow=7pt,
backgroundcolor=red!15,
innerleftmargin=5pt,
innerrightmargin=5pt,
innertopmargin=5pt,
leftmargin=0cm,
rightmargin=0cm,
innerbottommargin=5pt,
linewidth=1pt]{rBox}

\newmdenv[skipabove=7pt,
skipbelow=7pt,
backgroundcolor=blue2!25,
innerleftmargin=5pt,
innerrightmargin=5pt,
innertopmargin=5pt,
leftmargin=0cm,
rightmargin=0cm,
innerbottommargin=5pt,
linewidth=1pt]{dBox}
\newmdenv[skipabove=7pt,
skipbelow=7pt,
backgroundcolor=darkkblue!15,
innerleftmargin=5pt,
innerrightmargin=5pt,
innertopmargin=5pt,
leftmargin=0cm,
rightmargin=0cm,
innerbottommargin=5pt,
linewidth=1pt]{sBox}
\definecolor{darkblue}{RGB}{0,76,156}
\definecolor{darkkblue}{RGB}{0,0,153}
\definecolor{blue2}{RGB}{102,178,255}
\definecolor{darkred}{RGB}{195,0,0}

\newcommand{\nc}{\newcommand}
\nc{\rnc}{\renewcommand}
\nc{\lbar}[1]{\overline{#1}}
\nc{\ketbra}[2]{|#1\rangle\!\langle#2|}

\nc{\avg}[1]{\langle#1\rangle}
\nc{\smfrac}[2]{\mbox{$\frac{#1}{#2}$}}
\nc{\tr}{\operatorname{Tr}}
\nc{\ox}{\otimes}
\nc{\dg}{\dagger}
\nc{\dn}{\downarrow}
\nc{\cA}{{\cal A}}
\nc{\cB}{{\cal B}}
\nc{\cC}{{\cal C}}
\nc{\cD}{{\cal D}}
\nc{\cE}{{\cal E}}
\nc{\cF}{{\cal F}}
\nc{\cG}{{\cal G}}
\nc{\cH}{{\cal H}}
\nc{\cI}{{\cal I}}
\nc{\cJ}{{\cal J}}
\nc{\cK}{{\cal K}}
\nc{\cL}{{\cal L}}
\nc{\cM}{{\cal M}}
\nc{\cN}{{\cal N}}
\nc{\cO}{{\cal O}}
\nc{\cP}{{\cal P}}
\nc{\cQ}{{\cal Q}}
\nc{\cR}{{\cal R}}
\nc{\cS}{{\cal S}}
\nc{\cT}{{\cal T}}
\nc{\cU}{{\cal U}}
\nc{\cV}{{\cal V}}
\nc{\cX}{{\cal X}}
\nc{\cY}{{\cal Y}}
\nc{\cZ}{{\cal Z}}
\nc{\cW}{{\cal W}}
\nc{\csupp}{{\operatorname{csupp}}}
\nc{\qsupp}{{\operatorname{qsupp}}}
\nc{\var}{{\operatorname{var}}}
\nc{\rar}{\rightarrow}
\nc{\lrar}{\longrightarrow}
\nc{\polylog}{{\operatorname{polylog}}}
\nc{\wt}{{\operatorname{wt}}}
\nc{\av}[1]{{\left\langle {#1} \right\rangle}}
\nc{\supp}{{\operatorname{supp}}}

\nc{\argmin}{{\operatorname{argmin}}}

\def\i{\mathbf{i}}

\def\x{\xi}

\nc{\RR}{{{\mathbb R}}}
\nc{\CC}{{{\mathbb C}}}
\nc{\FF}{{{\mathbb F}}}
\nc{\NN}{{{\mathbb N}}}
\nc{\ZZ}{{{\mathbb Z}}}
\nc{\PP}{{{\mathbb P}}}
\nc{\QQ}{{{\mathbb Q}}}
\nc{\UU}{{{\mathbb U}}}
\nc{\EE}{{{\mathbb E}}}
\nc{\id}{{\operatorname{id}}}

\nc{\CHSH}{{\operatorname{CHSH}}}

\nc{\be}{\begin{equation}}
\nc{\ee}{{\end{equation}}}
\nc{\bea}{\begin{eqnarray}}
\nc{\eea}{\end{eqnarray}}
\nc{\<}{\langle}
\rnc{\>}{\rangle}
\nc{\rU}{\mbox{U}}

\nc{\ob}[1]{#1}

\nc{\SEP}{{\text{\rm SEP}}}
\nc{\NS}{{\text{\rm NS}}}
\nc{\LOCC}{{\text{\rm LOCC}}}
\nc{\PPT}{{\text{\rm PPT}}}
\nc{\EXT}{{\text{\rm EXT}}}
\nc{\Sym}{{\operatorname{Sym}}}

\nc{\ERLO}{{E_{\text{r,LO}}}}
\nc{\ERLOCC}{{E_{\text{r,LOCC}}}}
\nc{\ERPPT}{{E_{\text{r,PPT}}}}
\nc{\ERLOCCinfty}{{E^{\infty}_{\text{r,LOCC}}}}
\nc{\Aram}{{\operatorname{\sf A}}}

\usepackage{tikz}
\usepackage{hyperref}
\hypersetup{colorlinks=true,citecolor=blue,linkcolor=blue,filecolor=blue,urlcolor=blue,breaklinks=true}

\makeatletter
\def\grd@save@target#1{%
  \def\grd@target{#1}}
\def\grd@save@start#1{%
  \def\grd@start{#1}}
\tikzset{
  grid with coordinates/.style={
    to path={%
      \pgfextra{%
        \edef\grd@@target{(\tikztotarget)}%
        \tikz@scan@one@point\grd@save@target\grd@@target\relax
        \edef\grd@@start{(\tikztostart)}%
        \tikz@scan@one@point\grd@save@start\grd@@start\relax
        \draw[minor help lines,magenta] (\tikztostart) grid (\tikztotarget);
        \draw[major help lines] (\tikztostart) grid (\tikztotarget);
        \grd@start
        \pgfmathsetmacro{\grd@xa}{\the\pgf@x/1cm}
        \pgfmathsetmacro{\grd@ya}{\the\pgf@y/1cm}
        \grd@target
        \pgfmathsetmacro{\grd@xb}{\the\pgf@x/1cm}
        \pgfmathsetmacro{\grd@yb}{\the\pgf@y/1cm}
        \pgfmathsetmacro{\grd@xc}{\grd@xa + \pgfkeysvalueof{/tikz/grid with coordinates/major step}}
        \pgfmathsetmacro{\grd@yc}{\grd@ya + \pgfkeysvalueof{/tikz/grid with coordinates/major step}}
        \foreach \x in {\grd@xa,\grd@xc,...,\grd@xb}
        \node[anchor=north] at (\x,\grd@ya) {\pgfmathprintnumber{\x}};
        \foreach \y in {\grd@ya,\grd@yc,...,\grd@yb}
        \node[anchor=east] at (\grd@xa,\y) {\pgfmathprintnumber{\y}};
      }
    }
  },
  minor help lines/.style={
    help lines,
    step=\pgfkeysvalueof{/tikz/grid with coordinates/minor step}
  },
  major help lines/.style={
    help lines,
    line width=\pgfkeysvalueof{/tikz/grid with coordinates/major line width},
    step=\pgfkeysvalueof{/tikz/grid with coordinates/major step}
  },
  grid with coordinates/.cd,
  minor step/.initial=.2,
  major step/.initial=1,
  major line width/.initial=2pt,
}
\makeatother

\usepackage{thmtools}
\usepackage{thm-restate}
\usepackage{etoolbox}
\makeatletter
\def\problem@s{}
\newcounter{problems@cnt}

\newcommand{\allproblems}{\problem@s}
\makeatother

\begin{document}
\title{Efficient Inversion of Unknown Unitary Operations with Structured Hamiltonians}

\author{Yin Mo}
\affiliation{The Hong Kong University of Science and Technology (Guangzhou), Guangdong 511453, China}
\author{Tengxiang Lin}
\email{tengxianglin23@gmail.com}
\affiliation{The Hong Kong University of Science and Technology (Guangzhou), Guangdong 511453, China}
\author{Xin Wang}
\email{felixxinwang@hkust-gz.edu.cn}
\affiliation{The Hong Kong University of Science and Technology (Guangzhou), Guangdong 511453, China}

\begin{abstract}
Unknown unitary inversion is a fundamental primitive in quantum computing and physics. Although recent work has demonstrated that quantum algorithms can invert arbitrary unknown unitaries without accessing their classical descriptions, improving the efficiency of such protocols remains an open question. In this work, we present efficient quantum algorithms for inverting unitaries with specific Hamiltonian structures, achieving significant reductions in both ancilla qubit requirements and unitary query complexity. We identify cases where unitaries encoding exponentially many parameters can be inverted using only a single query. We further extend our framework to implement unitary complex conjugation and transposition operations, and develop modified protocols capable of inverting more general classes of Hamiltonians. We have also demonstrated the efficacy and robustness of our algorithms via numerical simulations under realistic noise conditions of superconducting quantum hardware. Our results establish more efficient protocols that improve the resources required for quantum unitary inversion when prior information about the quantum system is available, and provide practical methods for implementing these operations on near-term quantum devices.
\end{abstract}

\date{\today}

\maketitle

\section{Introduction} 
Governing by Schrodinger's equation, a closed quantum system evolves under unitary operation $U = e^{-iHt}$. 
Since the inverse of this evolution cannot be done directly as time doesn't go back, the simulation of reversing such a process with given oracle $U$ without its classical information is not only of interest to fundamental physics but also plays an important role in quantum information processing. 
Its applications include constructing crucial oracles within quantum algorithm frameworks~\cite{Gilyen2019,Wang2023,Odake2025a}, enabling quantum functional programming~\cite{Selinger2009,Odake2024a,Chiribella2009,Taranto2025}, and extracting properties of many-body physical systems or assisting Hamiltonian simulations~\cite{Leimkuhler2004,Lloyd1996,SchulteHerbruggen2006}. 

Due to the complexity of reversing unknown quantum processes, early studies primarily focused on approximate~\cite{Bisio2010,Quintino2022} and probabilistic unitary inversion~\cite{Quintino2019,Yang2021,Trillo2023}. 
In recent years have definitive results showed that quantum computer could reverse arbitrary-dimensional unitaries deterministically and exactly with finite queries of the gate itself~\cite{Yoshida2023,Chen2024}. Compared with tomography-based methods\cite{Baldwin2014,Gutoski2014,Mohseni2008,Haah2023}, which inevitably introduce errors with finite queries, the Quantum Unitary Reversal Algorithm (QURA)~\cite{Chen2024} provides acceleration in both time complexity and query complexity, which demonstrates fundamental advantages enabled by quantum computation. 

However, for large-qubit-number unitaries, QURA requires prohibitive resources, with query complexity growing exponentially. 
This may due to the fact that QURA is a general protocol which works for arbitrary unitary operations. 
Ref.~\cite{Odake2024} showed that this scaling is unavoidable, as the lower bound of the querying number is proven to be $d^2$ if all $d$-dimensional unitaries need to be reversed, making it impractical for large systems.

Noticing that 
the physical evolutions we considered or the quantum oracles we need to construct often possess special inherent structures. For instance, in Ising models, we know that the interactions only occur between adjacent qubits, e.g., $H=\sum a_{ij} Z_i Z_j + \sum b_i X_i$. 
A natural idea then arises: is it possible to construct more efficient deterministic and exact unitary inversion protocols, which are less general and only work for specific tasks or Hamiltonians? 
In this work, we answer this question affirmatively by showing several scenarios and conditions enabling efficient inversion, along with concrete implementation protocols. 
We also showed the robustness of our protocols when the condition doesn't fully satisfied and tested them under realistic noise conditions with IBM-Q cloud service. 
Additionally, we extend the ideas behind these protocols to efficiently implement unitary complex conjugation and transposition. 
Combining with the original QURA, it enables us to efficiently reverse unitaries with more general Hamiltonians. 
All these results show the possibility for making unitary inversion become more practical in specific scenarios.

To open the whole story, we want to highlight one important observation with a question: 
For an $N$-qubit unknown unitary gate, if we know its Hamiltonian can be decomposed into $M$ linearly independent terms while the coefficients before each term are unknown, what is the required query number for realizing its inversion? As a general $d$-dimensional unitary gate has $d^2$ linearly independent terms, which is exactly the lower bound of querying number derived in~\cite{Odake2024}, one might conjecture the answer be $O(M)$. Surprisingly we find that, for special Hamiltonian structures, the query number doesn't depend on $M$ but a quantity which we name as the `anti-commute set size', that will dramatically reduce the query complexity for reversing such unitary operations, and make quantum protocol essentially different from classical algorithms. 

\section{Setup}
Without loss of generality, 
we consider inverting an $N$-qubit unitary $U = e^{-iHt}$, where $H = \sum_{j=0}^{M-1} a_j P_j$ is decomposed in the Pauli basis. 
Denote $\mathcal{S} \coloneqq \{ P_j: 0\le j \le M-1 \}$ be the \textit{Pauli support} of $U$ and $H$, which is the set of all Pauli terms in $H$ (excluding identity terms, which only contribute to a global phase), this unitary is unknown in the sense that all coefficients $a_j$ and the evolution time $t$ are unknown, so there are $M$ independent changeable parameters.
For simplicity, in the following we use subscripts to denote each Pauli operators, e.g., $X_0 Z_2$ for a 3-qubit system means $X \otimes I \otimes Z$.

\section{Single-query unitary inverse without ancilla systems}
We first answer the previous question with a surprisingly easy protocol, that an $N$-qubit unitary with specific Hamiltonian structures could be reversed with querying it for only 1 time without ancilla qubits, while the changeable parameters could be exponential in $N$:

\begin{tcolorbox}[width=1.0\linewidth]
\begin{theorem}[Single--Query Inverse]\label{thm:1slot}
An $N$-qubit unknown unitary $U$ with Pauli support $\cS$ and $M$ independent changeable parameters $\{a_j\}$, can be reversed with 1 query without ancilla qubits, iff $\exists V$ s.t. $V$ anti-commutes with all \(P_j\) in \(\mathcal{S}\). $U^{\dag}$ can be realized deterministically and exactly by $V U V^{\dag}$, and  $M$ can be at most $2^{2N-1}$.
\end{theorem}
\end{tcolorbox}

\begin{sketch_of_proof}
While it may seem astonishing at first glance that an unknown unitary with exponentially many parameters can be inverted using just one query, the proof is straightforward.
Observing that $U^{-1}=e^{iHt}$ requires negating the coefficients of all Pauli terms in $H$, the property $V e^{-iHt} V^\dag = e^{-iV H V^\dag t}$ shows that when exist an operator $V$ anti-commutes with all elements in $\mathcal{S}$, $VHV^\dag$ transforms $H$ to $-H$, thereby implementing $U^{-1}$ directly. 
For the unknown parameter number, 
take $V=\bigotimes_{i=0}^{N-1} Z_i$ as an example, one could easily find that there exist $2^{2N} /2$ Pauli terms that anti-commute with $V$. 
The details of the proof is shown in Appendix. 
\end{sketch_of_proof}

As shown in \cref{subfig:chain}, the one-dimensional Ising model with $H=\sum a_{i} Z_i Z_{i+1} + \sum b_i X_i$ satisfies this condition. Choosing $V=(Z \otimes Y)^{\otimes N/2}$ enables implementing $U$ inverse via simply perform these local Pauli operations before and after $U$. 
Thus for such unitary, we don't need to know any parameter but could reverse it efficiently, and it is experimentally straightforward to realize.

\begin{figure}[H]
  \centering
  \scalebox{0.7}{%
    \begin{tabular}{cccc}
      \begin{subfigure}[b]{0.35\linewidth}
      \centering
      \resizebox{0.8\linewidth}{!}{%
        \begin{tikzpicture}[every node/.style={
                  circle,draw,fill=blue!20,
                  inner sep=0pt,minimum size=7pt}]
        \foreach \i in {1,...,6} {
          \node (P\i) at ({60*(\i-1)}:1.2) {};
        }
        \foreach \i in {1,...,5} {
          \pgfmathtruncatemacro{\j}{\i+1}
          \draw (P\i) -- (P\j);
        }
        \end{tikzpicture}
      }
      \caption{Linear chain (\(\checkmark\))}\label{subfig:chain}
      \end{subfigure}
      &
      \begin{subfigure}[b]{0.23\linewidth}
        \centering
        \resizebox{\linewidth}{!}{%
          \begin{tikzpicture}
            \foreach \x in {0,1,2,3}
              \foreach \y in {0,1,2,3}
                \node[circle,draw,fill=blue!20] (n\x\y) at (\x,\y) {};
            \foreach \x in {0,1,2}
              \foreach \y in {0,1,2,3}{
                \pgfmathtruncatemacro{\nx}{\x+1}
                \ifnum\y=1
                  \ifnum\x=2\else\draw (n\x\y)--(n\nx\y);\fi
                \else
                  \draw (n\x\y)--(n\nx\y);
                \fi
              }
            \foreach \x in {0,1,2,3}
              \foreach \y in {0,1,2}{
                \pgfmathtruncatemacro{\ny}{\y+1}
                \draw (n\x\y)--(n\x\ny);
              }
          \end{tikzpicture}
        }
        \caption{Grid (\(\checkmark\))}\label{subfig:grid}
      \end{subfigure}
      &
        \begin{subfigure}[b]{0.29\linewidth}
        \centering
        \resizebox{0.85\linewidth}{!}{%
          \begin{tikzpicture}
          \foreach \i in {1,...,7} {
            \node[circle,draw,fill=blue!20] (P\i) at ({360/7*(\i-1)+90}:1.5) {};
          }
          \foreach \i in {1,...,7} {
            \pgfmathtruncatemacro{\j}{mod(\i,7)+1}
            \draw (P\i) -- (P\j);
          }
          \end{tikzpicture}
        }
        \caption{Odd cycle (\(\times\))}\label{subfig:ring}
        \end{subfigure}
      &
      \begin{subfigure}[b]{0.28\linewidth}
        \centering
        \resizebox{\linewidth}{!}{%
          \begin{tikzpicture}
            \foreach \i/\ang in {1/90,2/18,3/306,4/234,5/162}{
              \coordinate (P\i) at (\ang:1.5);
            }
            \foreach \i in {1,...,5}{
              \foreach \j in {1,...,5}{
                \ifnum\i<\j
                  \draw (P\i) -- (P\j);
                \fi
              }
            }
            \foreach \i in {1,...,5}{
              \node[circle,draw,fill=blue!20] at (P\i) {};
            }
          \end{tikzpicture}
        }
        \caption{Pentagon (\(\times\))}\label{subfig:pentagon}
      \end{subfigure}
    \end{tabular}%
  }
  \caption{Examples of the Ising model. From Corollary~\ref{coro:1slot} one could directly see that (a) and (b) could be 1-query reversed as it has no cycle or only contains cycles with even number of nodes. 
  (c) and (d) could not be reversed as they contain cycles with odd number of nodes.}
  \label{fig:connectivity_examples}
\end{figure}
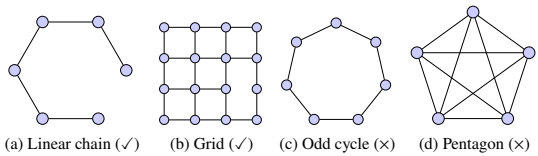

Although this result appears simple, it highlights a profound insight: For quantum tasks like unitary inversion, it is possible to bypass classical parameter estimation entirely through tailored quantum operations, dramatically reducing resource overhead. 
The most famous example of such idea that perform quantum task without classical information is Quantum Teleportation~\cite{Bennett1993}, which not only play important rules for quantum communication, but has also inspired fault-tolerant quantum computing nowadays. 
For unitary inversion, \cref{thm:1slot} shows that while the classical approach, which measure coefficients $\{a_j\}$ and $t$ via tomography, inevitably introduce errors with finite calls of measurements and face exponentially growing parameter numbers as $N$ increases, the circuit-based quantum approach may only require 1 query of the gate $U$ to realize the inversion perfectly.

After getting \cref{thm:1slot}, a natural question arises: Given a set $\mathcal{S}$, how can we determine whether there exists a $V$ anticommuting with all elements in $\mathcal{S}$, and could such $V$ be found efficiently? 
To address this, we give an equivalent condition of \cref{thm:1slot} so that whether $V$ exist can be checked easily. 
We then provide a Gaussian elimination-based algorithm to identify it.

\begin{tcolorbox}[width=1.0\linewidth]
\begin{corollary}[Condition for single--query inverse]\label{coro:1slot}
For an $N$-qubit unitary $U$ with Pauli support $\cS$,
$\exists V$ satisfying $V U V^\dag=U^\dag$ 
iff there is no subset with an odd number of elements in $\mathcal{S}$, i.e. \(\{P_{j_1},\ldots,P_{j_{2K+1}}\}\subseteq \mathcal{S}\), such that
$\prod_{k=1}^{2K+1} P_{j_k} \sim I$,
where “\(\sim\)” denotes equality up to a global phase.
\end{corollary}
\end{tcolorbox}

See the Appendix for detailed proof. 
Take the Ising model with $H=\sum a_{ij} Z_i Z_{j} + \sum b_i X_i$ as an example (see \cref{fig:connectivity_examples}), where edges connecting pairs of qubits represent $ZZ$ couplings. This criterion directly indicates which Hamiltonians admit inversion with a single query. Since the product of $ZZ$ couplings around a closed loop equals the identity, the structures in \cref{subfig:ring,subfig:pentagon} violate Corollary~\ref{coro:1slot}'s conditions and cannot be inverted in one query. Conversely, the structures in \cref{subfig:chain,subfig:grid} trivially satisfy the criterion and permit efficient inversion.

To obtain the anti-commuting Pauli operator $V$, we provide a Gaussian elimination-based algorithm. 
The core idea is to map $\{I,X,Y,Z\}$ to binary strings $\{00,01,11,10\}$ so that each $N$-qubit Pauli term is encoded as a $2N$-bit vector (e.g.\ $X_0Y_1$ maps to ``1101'', with the first $N$ bits for $X$ and the second $N$ bits for $Z$). Identifying operator $V$ reduces to solving a system of linear equations over $\mathbb{F}_2$: we seek a binary vector whose inner product modulo 2 with every vector in $\cS$ is one. 
The details of this algorithm is shown in Appendix, and we also provide a numerical program to calculate it~\cite{EfficientUnitaryInversion2025}. 
For a 10-qubit system with $M=10^5$ Pauli terms, the program could gets $V$ in less than one second.

\section{Multi-query unitary inverse without ancilla systems}
For Hamiltonians not satisfying \cref{thm:1slot}'s conditions, the core insight behind the theorem still enables us to find efficient unitary inversion protocols. Here, we show two scenarios that permit efficient inversion without ancilla qubits, which require multiple queries to the unitary.

The first scenario we considered is when all elements in $\mathcal{S}$ pairwise commute. 
Cluster model~\cite{Pachos2004}, e.g. $H = \sum_{i=0}^{N-3} a_i Z_i X_{i+1} Z_{i+2}$, is an example satisfying this condition, where the ground state of such system is the so-called cluster state playing an important role in measurement based quantum computing. 
Such a scenario also receives attention in phase transition study as it could raise all types of non-chiral topological phases of matter~\cite{Kastoryano2016,Bardet2023}.


\begin{tcolorbox}[width=1.0\linewidth]
\begin{theorem}\label{thm:commuting_comb}
For a unitary $U$ with elements in its Pauli support $\mathcal{S}$ pairwise commute, 
the anti-commute set $\mathcal{W}=\{V_0,V_1,...,V_{L-1}\}$ is a set such that for all $P_j \in \mathcal{S}$ there exist a Pauli operator $V_l \in \mathcal{W}$ satisfying $\{P_j, V_l\}=0$, and this unitary can be reversed with $2^L-1$ queries without ancilla qubits.
\end{theorem}
\end{tcolorbox}

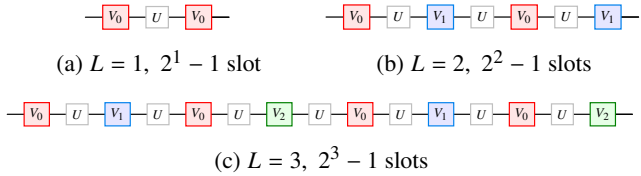
\begin{figure}[!htbp]
  \centering
  \begin{subfigure}[b]{0.5\columnwidth}
    \centering
    \makebox[\linewidth][c]{%
      \resizebox{0.48\columnwidth}{!}{%
        \begin{quantikz}[column sep=0.4cm]
          & \gate[style={draw=qred,text=qred,fill=red!10}]{V_0}
          & \gate[style={draw=gray!50,text=gray!50}]{U}
          & \gate[style={draw=qred,text=qred,fill=red!10}]{V_0}
          & \qw
        \end{quantikz}
      }%
    }
    \caption{$L=1,\;2^1-1$ slot}
    \label{fig:commuting_comb_L1} 
  \end{subfigure}%
  \hfill
  \begin{subfigure}[b]{0.5\columnwidth}
    \centering
    \makebox[\linewidth][c]{%
      \resizebox{\columnwidth}{!}{%
        \begin{quantikz}[column sep=0.4cm]
          & \gate[style={draw=qred,text=qred,fill=red!10}]{V_0}
          & \gate[style={draw=gray!50,text=gray!50}]{U}
          & \gate[style={draw=qblue,text=qblue,fill=blue!10}]{V_1}
          & \gate[style={draw=gray!50,text=gray!50}]{U}
          & \gate[style={draw=qred,text=qred,fill=red!10}]{V_0}
          & \gate[style={draw=gray!50,text=gray!50}]{U}
          & \gate[style={draw=qblue,text=qblue,fill=blue!10}]{V_1}
          & \qw
        \end{quantikz}
      }%
    }
    \caption{$L=2,\;2^2-1$ slots}
  \end{subfigure}

  \vspace{1ex}

  \begin{subfigure}[b]{0.98\columnwidth}
    \centering
    \makebox[\linewidth][c]{%
      \resizebox{\columnwidth}{!}{%
        \begin{quantikz}[column sep=0.4cm]
          & \gate[style={draw=qred,text=qred,fill=red!10}]{V_0}
          & \gate[style={draw=gray!50,text=gray!50}]{U}
          & \gate[style={draw=qblue,text=qblue,fill=blue!10}]{V_1}
          & \gate[style={draw=gray!50,text=gray!50}]{U}
          & \gate[style={draw=qred,text=qred,fill=red!10}]{V_0}
          & \gate[style={draw=gray!50,text=gray!50}]{U}
          & \gate[style={draw=green!50!black,text=green!50!black,fill=green!10}]{V_2}
          & \gate[style={draw=gray!50,text=gray!50}]{U}
          & \gate[style={draw=qred,text=qred,fill=red!10}]{V_0}
          & \gate[style={draw=gray!50,text=gray!50}]{U}
          & \gate[style={draw=qblue,text=qblue,fill=blue!10}]{V_1}
          & \gate[style={draw=gray!50,text=gray!50}]{U}
          & \gate[style={draw=qred,text=qred,fill=red!10}]{V_0}
          & \gate[style={draw=gray!50,text=gray!50}]{U}
          & \gate[style={draw=green!50!black,text=green!50!black,fill=green!10}]{V_2}
          & \qw
        \end{quantikz}
      }%
    }
    \caption{$L=3,\;2^3-1$ slots}
    \label{fig:commuting_comb_L3} 
  \end{subfigure}
  \caption{Unitary Inverse for condition in \cref{thm:commuting_comb}.}
  \label{fig:commuting_comb}
\end{figure}

Here we name the number of elements in set $\mathcal{W}$ as the `anti-commute set size' $|\mathcal{W}|=L$.
In \cref{fig:commuting_comb}, we present concrete quantum circuits for implementing unitary inversion when the anti-commute set size is 1 to 3. 
If $\mathcal{W}$ only contains 1 element, it reduces to the protocol of \cref{thm:1slot} as shown in \cref{sub@fig:commuting_comb_L1} (For Pauli operator $V^\dag = V$). 
The proof of how these circuits work and how to construct the circuit for large $L$ are both based on recurrence method, and are shown in detail in the Appendix. 

This protocol is already optimal as it achieves the lower bound derived in~\cite{Odake2024}, 
where authors showed that diagonal unitary inversion for $N$-qubit unitary requires $2^N-1$ queries. 
More generally, the query number in our protocol doesn't depend on $N$ but $L$, as not all commute terms need to appear in $\mathcal{S}$. 
Consider a 3-qubit Hamiltonian $H=\sum a_{ij} Y_i Y_j +\sum b_i Y_i$ with a cycle structure in \cref{sub@subfig:ring}, we can get its anti-commute set $\mathcal{W}=\{Z_0 Z_1, Z_1 Z_2\}$, then we can reverse this unitary with circuit shown in \cref{sub@fig:multislot_example}. It only needs to query this unknown unitary 3 times, achieving a drastic reduction from the general-case which requires 7 queries. 

The protocol in \cref{thm:commuting_comb} could then be generalized to contain anti-commute Pauli terms as stated in the following theorem:
\begin{tcolorbox}[width=1.0\linewidth]
\begin{theorem}\label{thm:commuting_comb2}
For a unitary $U$ with Pauli support $\cS$,
if $\mathcal{S}$ can be decomposed into two subsets $\hat{\mathcal{S}}_0$ and $\hat{\mathcal{S}}_1$ that commute with each other, 
where there exist a Pauli operator $V_0$ anti-commute with $\hat{\mathcal{S}}_1$, all elements in $\hat{\mathcal{S}}_0$ pairwise commute and can be reversed based on the anti-commute set $\mathcal{W}=\{V_1,...,V_{L-1}\}$, then this unitary can be reversed without ancilla qubits by querying $U$ for $2^L-1$ times. 
\end{theorem}
\end{tcolorbox}
The details of the proof is shown in Appendix, while the core idea for this protocol is based on the identity that when $A$ commute with $B$, $e^{A+B} e^{A-B} = e^{2A}$. 
Noticing that if all elements in $\hat{\mathcal{S}}_1$ pairwise commute, \cref{thm:commuting_comb2} reduced to \cref{thm:commuting_comb} that is why it can be regarded as a generalization. 
From the above protocols, one could see that the query number for inversion doesn't depend on the qubit number $N$ but on the `anti-commute set size'.

\begin{figure}[H]
  \centering
  \begin{subfigure}[b]{0.48\linewidth}
    \centering
    \resizebox{\linewidth}{!}{%
      \begin{quantikz}[column sep=0.4cm]
        & \qw
        & \gate[style={draw=qred,text=qred,fill=red!10}]{Z}
        & \gate[3,style={draw=gray!50,text=gray!50}]{U}
        &
        & \gate[3,style={draw=gray!50,text=gray!50}]{U}
        & \gate[style={draw=qred,text=qred,fill=red!10}]{Z}
        & \gate[3,style={draw=gray!50,text=gray!50}]{U}
        &
        & \qw \\
        & \qw
        & \gate[style={draw=qred,text=qred,fill=red!10}]{Z}
        &
        & \gate[style={draw=qblue,text=qblue,fill=blue!10}]{Z}
        &
        & \gate[style={draw=qred,text=qred,fill=red!10}]{Z}
        &
        & \gate[style={draw=qblue,text=qblue,fill=blue!10}]{Z}
        & \qw \\
        &     &     &
        & \gate[style={draw=qblue,text=qblue,fill=blue!10}]{Z}
        &     &     &
        & \gate[style={draw=qblue,text=qblue,fill=blue!10}]{Z}
        & \qw
      \end{quantikz}
    }
    \caption{}
    \label{fig:multislot_example}
  \end{subfigure}%
  \hfill
  \begin{subfigure}[b]{0.48\linewidth}
    \centering
    \resizebox{\linewidth}{!}{%
      \begin{quantikz}[column sep=0.4cm]
        & \qw
        & \gate[3,style={draw=gray!50,text=gray!50}]{U}
        & \gate[style={draw=qblue,text=qblue,fill=blue!10}]{Y}
        & \gate[3,style={draw=gray!50,text=gray!50}]{U}
        &
        & \gate[3,style={draw=gray!50,text=gray!50}]{U}
        & \gate[style={draw=qblue,text=qblue,fill=blue!10}]{Y}
        & \qw \\
        & \gate[style={draw=qred,text=qred,fill=red!10}]{Y}
        &
        & \gate[style={draw=qblue,text=qblue,fill=blue!10}]{X}
        &
        & \gate[style={draw=qred,text=qred,fill=red!10}]{Y}
        &
        & \gate[style={draw=qblue,text=qblue,fill=blue!10}]{X}
        & \qw \\
        & \qw
        &
        & \gate[style={draw=qblue,text=qblue,fill=blue!10}]{Z}
        &
        & \qw
        &
        & \gate[style={draw=qblue,text=qblue,fill=blue!10}]{Z}
        & \qw
      \end{quantikz}
    }
    \caption{}
    \label{fig:commuting_example}
  \end{subfigure}
  \caption{Inverse circuit for $H=\sum a_{ij} Y_i Y_j +\sum b_i Y_i$ and $H = a Z_0 X_1 Z_2 + \sum_{i=0}^{1} b_i X_i X_{i+1} + \sum_{i=0}^{2} c_i X_i$.}
  \label{fig:anti-commuting_comb}
\end{figure}
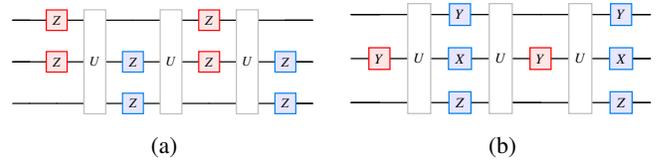

Cluster-Ising model~\cite{Smacchia2011} of 3-qubit is an example satisfying this condition with $H = a Z_0 X_1 Z_2 + \sum_{i=0}^{1} b_i X_i X_{i+1} + \sum_{i=0}^{2} c_i X_i$. 
Here we can get subset $\hat{S}_0=\{X_1\}$, with $V_0 = Y_1, V_1 = Y_0X_1Z_2$. Thus the circuit which reverses such unitary is shown in \cref{fig:commuting_example}. 
Another example is the odd-cycle Ising model in \cref{subfig:ring}, which cannot be reversed with 1 query, but if two Pauli $X$ terms are missing as $H=\sum_{i=0}^{6} a_{i} Z_i Z_{i+1} +\sum_{i=1}^{5} b_i X_i$, then it can be reversed by choosing $\hat{S}_0=\{Z_6 Z_0\}$ based on \cref{thm:commuting_comb2}. 
We give the detailed discussion for this and a more difficult example which require 7 queries for inversion in the Appendix.

\section[Generalization to U* and UT]{Generalization to $U^*$ and $U^T$} 
The idea behind realizing unitary inverse could also be generalized to realize unitary complex conjugate $U^*$ and unitary transpose $U^T$. 
For complex conjugate, it is well known that $YUY = U^*$ for qubit, but it cannot be extended to higher dimensions directly. The only solution for realizing $U^*$ for $d$-dimensional unitary previously is by using $d-1$ gates parallelly, which requires $d-2$ qudits as the ancilla system~\cite{Miyazaki2019}. 
Thus, for $N$-qubit system, the protocol in this work will significantly reduce the ancilla qubit number from $2^N-1$ to 0. 
We show the details of how to generalize \cref{thm:1slot,thm:commuting_comb,thm:commuting_comb2} to  $U^*$ and $U^T$ in the Appendix. 
Here we give the generalization of Corollary~\ref{coro:1slot} to show when $U^*$ can be realized with one query of $U$, from which one could see that single-qubit unitaries can be transformed by $YUY=U^*$ can be regarded as a special example of it:
\begin{tcolorbox}[width=1.0\linewidth]
\begin{theorem}[Single--query complex conjugate]\label{coro:1slot-complex}
For an $N$-qubit unitary $U$ with Pauli support $\cS$, it can be transformed to its complex conjugate with a Pauli operator $V$ such that $U^{*} = V U V$, iff for an arbitrary subset \(\hat{\mathcal S}\subseteq \mathcal{S}\), if $\prod_{i\in \hat{\mathcal S}} P_{i} \sim I$, there are even numbers of Pauli operators in \(\hat{\mathcal S}\) which has even $Y$ terms. 
\end{theorem}
\end{tcolorbox}

One observation is important to point out here: 
Traditionally people always assume that implementing the $U^\ast$ of an unknown unitary is inherently simpler than implementing its inverse. However, by rigorously comparing the Hamiltonian conditions required for 1-query realizing $U^\dag$ and 1-query realizing $U^\ast$, we could find that this is not universally true. Specifically, there exist Hamiltonian structures where $U^\dag$ can be implemented with a single query, while $U^\ast$ cannot. Note, this phenomenon also exist for unitary transposition, where $U^T$ can be implemented with a single query, while $U^\ast$ and $U^T$ cannot. These findings reveal that the relative difficulty of unitary transformations, inverse, conjugate, and transpose, is highly related to the structure of $H$, fundamentally overturning prior intuition about their inherent complexities.

\section{Modified QURA for more general Hamiltonians}

For Hamiltonians not satisfying the aforementioned conditions, we propose a modified QURA-based protocol to realize its inverse. 
We first recap that QURA is realized based on two important subcircuits, where the first one is used to deterministically and exactly realize $U^*$, while the seconde one is to use a subcircuit with linear combination of unitaries (LCU) to perform $U^T$ so that its amplitude-amplification (AA) process could work. 
If $U^*$ can be realized more efficiently and $U^T$ has a simpler decomposition, resource overhead could be significantly reduced.
As an example if $H=a X_0 X_1 X_2 + bY_0 Y_1 Y_2 + cZ_0 Z_1 Z_2$, the original QURA~\cite{Chen2024} for 3-qubit unitary needs to use 25 ancilla qubits and query gate $U$ for 103 times, 
while with modified QURA, we only need to use 3 ancilla qubits with querying $U$ for 5 times to realize its inversion. 

For realizing $U^*$, besides the protocol mentioned above, through numerical optimization based on PQComb~\cite{Mo2025}, we also find that there exist 2-qubit Hamiltonians which can be transformed by sequentially using gate $U$ for 3 times with ancilla qubits less than four. 
As an example, Hamiltonian with support $\mathcal{S} = [X_0, Y_0, Z_0, X_1, X_0X_1, Z_0X_1]$ can be transformed into its complex conjugate with 2 ancilla qubits, while non-Pauli operations are required. 
Such an example also shows the trade-off that increasing circuit depth, we could reduce ancilla overhead, which is maybe useful for near-term applications. 
More examples and the program are shown in the Appendix.

For unitary transpose, to realize $U^T$ by LCU circuit, we need to get a linear decompositions of $U$ with some specific conditions. For qubits, QURA uses $IUI+XUX-YUY+ZUZ=2U^T$ to construct $U^T$, which holds for arbitrary qubit unitary, it leads to run AA for 3 rounds(include the initialization). Specific Hamiltonians permit dramatic simplifications, e.g. $XUX=U^T$ holds for $H=aX+bY$, it reduces the round of AA to just 1. 
Due to the special condition for this problem, identifying such decompositions for a general $N$-qubit Pauli support $\mathcal{S}$ is a very hard problem. 
To deal with this, we propose a parameterized optimization method based on PQComb~\cite{Mo2025}, which transform the problem into optimizing success probability using 1-query of gate $U$ to realize $=U^T$ exactly. 
The details of this technique is shown in Appendix, that we believe it may also inspire quantum methods to solve difficult optimization problems.
Our numerical results for 3-qubit Hamiltonian with 6-terms in $\mathcal{S}$ shows that about $65\%$ unitaries only need 1 round of AA, and  $10\%$ need 3 rounds, significantly reduced from the original QURA which needs 13 rounds of AA for 3 qubit unitaries.

\section{The robustness of the unitary inversion protocols}
We numerically tested the robustness of the unitary inverse protocols and found that even if the condition doesn't fully satisfied, the error is still very small. 
We tested by inserting new unitaries into the original circuit with  $H = a \mathcal{S} + b \mathcal{S}'$, where $\mathcal{S}'$ be the set of Pauli terms not in $\mathcal{S}$, and $|b|=\delta|a|$. 
Using the ideal inverse circuit for $\mathcal{S}$, the error of the average fidelity is $O(\delta^2)$ for all examples in \cref{fig:commuting_comb,fig:anti-commuting_comb}, where the details of the program and the numerical results are shown in the Appendix. 
This result showed that even if a small amount of Hamiltonian is not in the support $\mathcal{S}$ our protocol is still a good approximate inversion protocol, and may also inspire hybrid protocols, with small amount of measurement used to first identify $U$'s structure and construct more efficient inversion circuit later. 

\begin{figure}[htbp]
  \centering
  \includegraphics[width=\linewidth]{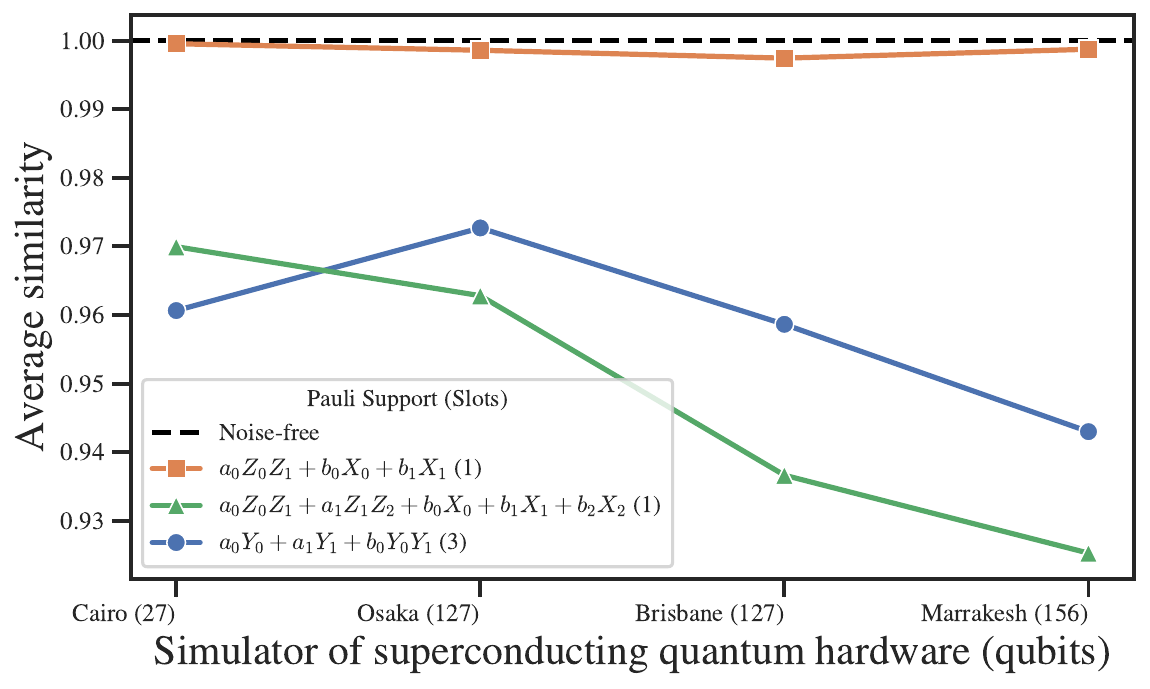}
  \caption{Simulation of the inverse protocols under the noise settings of four real quantum devices. 
  The ``slots'' for each Pauli support indicate the number of queries required to implement the inverse protocol, while the ``qubits'' for each device indicate the number of qubits available on the device.}
  \label{fig:IBM_noise_backend}
\end{figure}

We have also simulated the performance of the circuit under realistic noise conditions by utilizing the IBM-Q cloud service, with noise settings from four different IBM quantum devices. In the experiments, we chose 2-qubit and 3-qubit Ising models as examples where inversion can be achieved with a single query, and the 2-qubit Hamiltonian $H= a_{0} Y_0 Y_1 +\sum_{i=0}^{1} b_i Y_i$ as an example requiring three queries to implement the inverse. 
As shown in \cref{fig:IBM_noise_backend}, the numerical simulations of our protocols for 2-qubit and 3-qubit quantum systems all achieve high-accuracy performance, while in~\cite{Mo2025} the single-qubit general unitary inverse circuit is also simulated, with the average fidelity falling below 0.9 across all simulators. This comparison highlights the practicality of our protocol and the advantage of designing unitary inverse circuits for quantum dynamics with certain Hamiltonian structures, which has significant practical value for near-term quantum hardware.

\section{Discussion} 
In this work, we showed that for $N$-qubit unitaries with specific Hamiltonian structures, inversion could be realized with substantially improved efficiency. Compared with the general protocol, the cost, including the number of ancilla qubits and the number of queries to the unitary gate, is significantly reduced. We derived the detailed conditions for such Hamiltonians and provided the protocols for realizing the inversion. We have also generalized these protocols to transform unknown unitaries to their complex conjugates and transposes, which are also important for quantum computing. For instance,~\cite{King2024} showed that access to both a quantum state and its complex conjugate enables exponential speedup in learning certain properties. Integrating our protocols could thus be used to extract specific row information from $U$. We demonstrated the efficacy and robustness of our algorithms via numerical simulations under realistic noise conditions of superconducting quantum hardware.

For most protocols we provided, it can be seen that ancilla qubits are not required and the operations that need to be performed are local Pauli gates, which makes them practical for near-term devices. In this work, we showed the inverse protocols for Ising models with different structures and the Cluster-Ising model. It would be interesting to study other Hamiltonian models in the future to see whether they could also be reversed without ancilla qubits, or what the cost would be when reversing with modified QURA.

An observation worth highlighting is that \cref{thm:1slot} essentially shows the difference between quantum computing and its classical counterpart. For a task that classically requires measuring an exponential number of parameters, a quantum approach could achieve it with just a single query. This phenomenon may remind us of the Deutsch-Jozsa algorithm~\cite{Deutsch1992}, which determines whether a function is constant or not with just one query to a quantum oracle. Since in our task, the quantum protocol doesn't need to know the information of any parameters, we also expect it may provide ideas for quantum cryptography.

Besides, through designed robustness testing of our method, we showed that even if a small portion of the Hamiltonian is not in the support, the protocol still works with high precision. This may inspire studies combining recent works on Hamiltonian structure learning~\cite{Bakshi2024}, which tries to determine the Pauli support of an unknown unitary. Through this combination, we may develop hybrid protocols that first identify $U$'s structure and then realize its inverse efficiently. In all, our results reveal broader possibilities for unitary transformation and functional programming~\cite{Taranto2025}, and may shed light on quantum algorithm design.


\textbf{Acknowledgement.---} The authors would like to thank Yu-Ao Chen for helpful discussions and the useful comments given by reviewers from AQIS2025. The codes for this paper are available on~\cite{EfficientUnitaryInversion2025} and the correspondence regarding the codes should be directed to T. Lin. This works was partially supported by the National Key R\&D Program of China (Grant No.~2024YFB4504004), the National Natural Science Foundation of China (Grant. No.~12447107), the Guangdong Provincial Quantum Science Strategic Initiative (Grant No.~GDZX2403008, GDZX2403001), the Guangdong Provincial Key Lab of Integrated Communication, Sensing and Computation for Ubiquitous Internet of Things (Grant No. 2023B1212010007), the Quantum Science Center of Guangdong-Hong Kong-Macao Greater Bay Area, and the Education Bureau of Guangzhou Municipality.

\mathcode`\:=\string"8000
\def\vcentcolon{\mathrel{\mathop\ordinarycolon}}
\begingroup \catcode`\:=\active
  \lowercase{\endgroup
  \let :\vcentcolon
  }

\definecolor{darkblue}{RGB}{0,76,156}
\definecolor{darkkblue}{RGB}{0,0,153}
\definecolor{blue2}{RGB}{102,178,255}
\definecolor{darkred}{RGB}{195,0,0}

\captionsetup{justification=raggedright, singlelinecheck=false}
 
\definecolor{colortwo}{rgb}{0.4,0.77,0.17}
\definecolor{colorthree}{rgb}{0.01,0.51,0.93}

\newenvironment{changemargin}[2]{%
\begin{list}{}{%
\setlength{\topsep}{0pt}%
\setlength{\leftmargin}{#1}%
\setlength{\rightmargin}{#2}%
\setlength{\listparindent}{\parindent}%
\setlength{\itemindent}{\parindent}%
\setlength{\parsep}{\parskip}%
}
\item[]}{\end{list}}






\appendix

\vspace{3cm}
\onecolumngrid
\vspace{2cm}

\clearpage
\begin{center}
\Large{\textbf{Appendix for Efficient Inversion of Unknown Unitary Operations with Structured Hamiltonians}}
\end{center}


\setcounter{subsection}{0}
\setcounter{table}{0}
\setcounter{figure}{0}
\setcounter{equation}{0}
\setcounter{table}{0}
\setcounter{section}{0}
\setcounter{figure}{0}
\numberwithin{equation}{section}
\renewcommand{\theproposition}{S\arabic{proposition}}
\renewcommand{\thedefinition}{S\arabic{definition}}
\renewcommand{\thefigure}{S\arabic{figure}}
\setcounter{equation}{0}
\setcounter{table}{0}
\setcounter{section}{0}
\setcounter{figure}{0}

In the appendix, we mainly give the proofs for the Theorems. 
Here we first recap the basic settings again: 
We consider inverting an $N$-qubit unitary $U = e^{-iHt}$, where $H = \sum a_i P_i$ is decomposed in the Pauli basis.
For clarity, an $N$-qubit Pauli operator can be denoted using subscripts to represent each Pauli operators on specific qubits, e.g., $X_0 Z_2$ for a 3-qubit system means $X \otimes I \otimes Z$. 
Denote $\mathcal{S} \coloneqq \{ P_i: 0\le i \le M-1 \} $ be the set of all Pauli terms in $H$ which contains $M$ elements (excluding identity terms, which only contribute to a global phase).  We define $\cS$ as the Pauli support of $U$ and $H$.
To reverse this unknown unitary, we assume $S$ is given, while the coefficients $a_i$ and the evolution time $t$ are unknown. 

In Appendix~\ref{appendix:inverse}, we present detailed proofs for Theorems of unitary inverse in the main text, and the details of the Gaussian elimination method to find the anti-commute set. 
In Appendix~\ref{appendix:robustness}, we show the numerical results to test the robustness of these protocols. 
In Appendix~\ref{appendix:conjugate}, we generalize these theorems to transform unknown unitaries to its complex conjugate, and in Appendix~\ref{appendix:PQComb_conjugate} we show how we use PQComb find some other protocols for unitary complex conjugate. 
Finally, in Appendix~\ref{appendix:transpose} we discuss the conditions for linearly decomposing unitary transpose to realize modified Quantum Unitary Reversal Algorithm(QURA), and provide a parameterized optimization method to get the linear decomposition. 

$\qquad$

 
\setcounter{theorem}{0}
\setcounter{definition}{0}
\renewcommand{\thetheorem}{A\arabic{theorem}}
\renewcommand{\theproposition}{A\arabic{proposition}}
\renewcommand{\thecorollary}{A\arabic{corollary}}
\section{Proofs of Theorems for Unitary Inversion}~\label{appendix:inverse}
\renewcommand\theproposition{1}
\begin{theorem}[Single-Query Inverse]\label{thm:1}
An $N$-qubit unknown unitary $U$ with Pauli support $\cS$ and $M$ independent changeable parameters $\{a_j\}$, can be reversed with 1 query without ancilla qubits, iff $\exists V$ s.t. $V$ anti-commutes with all \(P_j\) in \(\mathcal{S}\). $U^{\dag}$ can be realized deterministically and exactly by $V U V^{\dag}$, and $M$ can be at most $2^{2N-1}$.
\end{theorem}
\renewcommand{\theproposition}{S\arabic{proposition}}
\begin{proof}
Substitute $U$ with $e^{-it\sum a_j P_j}$, we can get $U^\dag = e^{it\sum a_j P_j}$. As for an arbitrary unitary $V$ we have $V e^{-iHt} V^\dag = e^{-i VHV^\dag t}$, $VUV^\dag = U^\dag$ is equivalent as:
\begin{align}
    e^{-it\sum -a_j P_j} = e^{-it\sum a_j V P_j V^\dag} \, .
\end{align}
As this equality holds for arbitrary coefficients $a_j$, we get 
\begin{align}
    V P_j V^\dag = -P_j \, , \forall P_j \in S \, ,
\end{align}
that is $V$ anti-commutes with all \(P_j\) in \(\mathcal{S}\): \(\{V,P_j\}=0\), \(\forall j\).

To see that $M$ can be at most $2^{2N-1}$, we first show that if $U$ can be reversed with 1 query, we can always choose $V$ to be a Pauli operator, i.e. $V = \otimes_{i=0}^{N-1} \sigma_i$, where $\sigma_i$ is the Pauli operator on the $i$-th qubit. 
Suppose there exists $V$ such that $\{V,P_j\}=0$, $\forall P_j \in S$, then we can decompose $V$ in Pauli basis as $V = \sum_{k=0}^{4^N-1} a_k' P_k'$. 
Using the property that for two Pauli operators, $P_j P_k P_j$ either equal to $P_k$ or $-P_k$, we can get that 
\begin{align}
    \nonumber\{V,P_j\}=0 &\iff P_j V P_j = -V \\
    &\iff P_j P_k' P_j = -P_k' \, \qquad \, \forall a_k' \neq 0 \, ,
\end{align}
which means that there exist $P_k^{'}$ anti-commute with $S$. 
Then it is easy to see that $M$ is at most $2^{2N-1}$ as for arbitrary $N$-qubit Pauli operator $P$, there are $2^{2N-1}$ different Pauli terms anti-commute with it.
\end{proof}

From now on, to find $V$ anti-commute with set $S$, we only consider $V$ to be a Pauli operator, thus $V^\dag = V$.

\renewcommand\theproposition{2}
\begin{corollary}\label{coro:2}
For an $N$-qubit unitary $U$ with Pauli support $\cS$,
$\exists V$ satisfying $V U V^\dag=U^\dag$ 
iff there is no subset with an odd number of elements in $\mathcal{S}$, i.e. \(\{P_{j_1},\ldots,P_{j_{2K+1}}\}\subseteq \mathcal{S}\), such that
$\prod_{k=1}^{2K+1} P_{j_k} \sim I$,
where “\(\sim\)” denotes equality up to a global phase. (i.e. the product is proportional to the identity).
\end{corollary}
\renewcommand{\theproposition}{S\arabic{proposition}}
\begin{proof}
We first set $V$ to be a Pauli operator $P$. Suppose $P$ anti-commute with $P_i$ and $P_j$, which means $P P_i P = -P_i$ and $P P_j P = -P_j$. Then we can get 
\begin{align}
    \nonumber P P_i P_j P &= P P_i P P P_j P \\
    \nonumber &= \Big(-P_i \Big)  \Big(-P_j \Big) \\
    &= P_i P_j \, .
\end{align}
It shows that $P_i$, $P_j$ and $P_i P_j$ cannot be in $S$ simultaneously. 
This result can be easily generalized to show that if $\prod_{k=1}^{2K+1} P_{j_k} \sim I$, these Pauli operators cannot be reversed simultaneously. 
\end{proof}

$\qquad$

\noindent\textbf{Pauli mapping and $\mathbb{F}_2$ elimination.}
It is not difficult to verify that when no such subset exist, we can easily find a Pauli operator $V$ anti-commute with $\mathcal{S}$ base on Gaussian elimination over \(\mathbb{F}_2\). 

We first map each \(N\)-qubit Pauli operator \(P_j\) in \(\mathcal{S}\) into a binary vector of length \(2N\): the first \(N\) bits indicate \(X\) components, the next \(N\) bits indicate \(Z\) components (\(Y\) corresponds to both \(X\) and \(Z\) at that position). 
As an example a 3-qubit Pauli operator $X_0Y_2$ will be mapped to $101001$ (the first 1 corresponds to $X_0$, the third and sixth 1 correspond to $Y_2$). 
Collecting these as rows forms a binary matrix \(A\). 
The anti-commutation condition between \(P_j\) (vector \(\mathbf{p}_j\)) and Pauli \(V\) (vector \(\mathbf{v}\)) is given by the symplectic inner product:
\[
\langle \mathbf{p}_j, \mathbf{v} \rangle_s = \mathbf{p}_j^X \cdot \mathbf{v}^Z + \mathbf{p}_j^Z \cdot \mathbf{v}^X \mod 2 = 1,
\]
where \(\mathbf{p}_j^X\) and \(\mathbf{p}_j^Z\) are the \(X\) and \(Z\) parts of \(\mathbf{p}_j\) respectively.

To find $V$ such that this holds for all \(P_j\), we construct the augmented matrix \([A \mid \mathbf{1}]\) over \(\mathbb{F}_2\), where each row corresponds to the symplectic inner product constraint for a \(P_j\), and \(\mathbf{1}\) is a column vector of ones. 
Then the solution \(\mathbf{v}\) can be found by Gaussian elimination, and the code is shown in~\cite{EfficientUnitaryInversion2025}.  
If a single \(\mathbf{v}\) suffices, the solution is found efficiently (e.g., for \(N=10\) and \(|\mathcal{S}|=10^5\), in less than one second). 

$\qquad$

Now we give the proof for \cref{thm:commuting_comb,thm:commuting_comb2} in the main text.

\renewcommand\theproposition{3}
\begin{theorem}\label{thm:3}
For a unitary $U$ with elements in its Pauli support $\mathcal{S}$ pairwise commute, the anti-commute set $\mathcal{W}=\{V_0,V_1,...,V_{L-1}\}$ is a set such that for all $P_j \in \mathcal{S}$ there exist a Pauli operator $V_l \in \mathcal{W}$ satisfying $\{P_j, V_l\}=0$, and this unitary can be reversed with $2^L-1$ queries without ancilla qubits.
\end{theorem}
\renewcommand{\theproposition}{S\arabic{proposition}}
\begin{proof}
We first give the circuit for reversing such a unitary gate with querying $U$ for $2^L-1$ times by the recurrence method:
\begin{itemize}
    \item When $\mathcal{W}$ only contains one term, the whole circuit is $U^\dag = V_0 U V_0$, where we denote $f_1(U)$ be the circuit without the first term $V_0$ as $f_1(U) = U V_0$.
    \item When $\mathcal{W}$ contains $L$ terms, and the circuit $f_{L-1}(U)$ has been constructed, the whole circuit is $U^\dag =  V_{L-1} f_{L-1}(U) U V_{L-1} f_{L-1}(U)$, and we denote $f_{L}(U)$ be the circuit without the first term $V_{L-1}$ as $f_{L}(U) = f_{L-1}(U) U V_{L-1} f_{L-1}(U)$. In all, this circuit queries $U$ for $2^L-1$ times.
\end{itemize}
The concrete quantum circuits for implementing unitary inversion when the anti-commute set $\mathcal{W}$ contains 1-3 elements are shown as \cref{fig:commuting_comb} in the main text.

Now we show how this circuit works. 
Denote the Pauli operators in \(\mathcal{S}\) which anti-commute at least one term in $\{V_0,V_1,...,V_{L-2}\}$ as $\mathcal{S}_{L-1}$, and the remaining terms as \(\tilde{\mathcal S}_{L-1}\), we can get that $\tilde{\mathcal S}_{L-1}$ commute with $\{V_0,V_1,...,V_{L-2}\}$ and anti-commute with $V_{L-1}$. 
Denote $U_{L-1} = e^{-it(\sum_{j \in \mathcal{S}_{L-1}} a_j P_j)}$ and $\tilde{U}_{L-1} = e^{-it(\sum_{j \in \tilde{\mathcal S}_{L-1}} a_j P_j)}$. As all elements in \(\mathcal{S}\) pairwise commute, $U = U_{L-1}\cdot\tilde{U}_{L-1}$. 
As $\tilde{\mathcal S}_{L-1}$ commute with $\{V_0,V_1,...,V_{L-2}\}$ and $\mathcal{S}_{L-1}$, we can get that $V_{L-2}f_{L-1}(U) = \tilde{U}_{L-1}^{(2^{L-1}-1)} \cdot U_{L-1}^\dag$ ($\tilde{U}_{L-1}$ to the power $(2^{L-1}-1)$). Then we can get: 
\begin{align}
    \nonumber V_{L-1} f_{L-1}(U) U V_{L-1} f_{L-1}(U) &=  V_{L-2}V_{L-1} U \tilde{U}_{L-1}^{(2^{L-1}-1)} \cdot U_{L-1}^\dag V_{L-2}V_{L-1} \tilde{U}_{L-1}^{(2^{L-1}-1)} U_{L-1}^\dag\\
    \nonumber &\sim  V_{L-2} V_{L-1} \tilde{U}_{L-1}^{(2^{L-1})} V_{L-1} V_{L-2} U_{L-1}^\dag \tilde{U}_{L-1}^{(2^{L-1}-1)}\\
    \nonumber &= V_{L-2}V_{L-2} U_{L-1}^\dag \tilde{U}_{L-1}^{(2^{L-1}-1)} \tilde{U}_{L-1}^{(2^{L-1})\dag}  \\
    \nonumber &= U_{L-1}^\dag \tilde{U}_{L-1}^\dag \\
    &= U^\dag
\end{align}
\end{proof}

To get the anti-commute set $\mathcal{W}$, we could use $\mathbb{F}_2$ elimination recursively, while we should point out that how to find the smallest set is a difficult problem, and is interesting for further study. 
In the numerical program we provide, we seek \(V\) such that this holds for as many \(P_j\) as possible.
We still first construct the augmented matrix \([A \mid \mathbf{1}]\) over \(\mathbb{F}_2\). After finding \(V\) based on Gaussian elimination, we remove all \(P_i\) for which the constraint is satisfied, and recursively repeat the process on the remaining rows (ignoring any contradictory rows that cannot be satisfied). This recursive elimination yields a set \(\mathcal{W} = \{V_1, V_2, \ldots\}\) such that every \(P_i\) anti-commutes with at least one \(V_j\).

\renewcommand\theproposition{4}
\begin{theorem}\label{thm:4}
For a unitary $U$ with Pauli support $\cS$, if $\mathcal{S}$ can be decomposed into two subsets $\hat{\mathcal{S}}_0$ and $\hat{\mathcal{S}}_1$ that commute with each other, where there exist a Pauli operator $V_0$ anti-commute with $\hat{\mathcal{S}}_1$, all elements in $\hat{\mathcal{S}}_0$ pairwise commute and can be reversed based on the anti-commute set $\mathcal{W}=\{V_1,...,V_{L-1}\}$, then this unitary can be reversed without ancilla qubits as querying $U$ for $2^L-1$ times. 
\end{theorem}
\renewcommand{\theproposition}{S\arabic{proposition}}
\begin{proof}
The key point for the proof is by using the equality that when matrix $A$ and $B$ commute, $e^{A+B} \cdot e^{A-B} = e^{2A}$, and the circuit is constructed recursively almost the same as in \cref{thm:3}: 
\begin{itemize}
    \item When $\hat{\mathcal{S}}_0$ has no elements, the whole circuit is $U^\dag = V_0 U V_0$, where we denote $f_1(U)$ be the circuit without the first term $V_0$ as $f_1(U) = U V_0$.
    \item When $\mathcal{W}$ contains $L-1$ terms, and the circuit $f_{L-1}(U)$ has been constructed, the whole circuit is $U^\dag =  V_{L-1} f_{L-1}(U) U V_{L-1} f_{L-1}(U)$, and we denote $f_{L}(U)$ be the circuit without the first term $V_{L-1}$ as $f_{L}(U) = f_{L-1}(U) U V_{L-1} f_{L-1}(U)$. In all, this circuit queries $U$ for $2^L-1$ times.
\end{itemize}

To see this circuit works, here we first decompose $U$ as $U=U_0 \cdot U_1 = e^{-it(\sum_{j \in \mathcal{\hat{S}}_0} a_j P_j)} \cdot e^{-it(\sum_{j \in \mathcal{\hat{S}}_1} a_j P_j)}$, then one important thing is to notice that 
\begin{align}
    V_0 U V_0 = U_0 \cdot U_1^\dag = e^{-it(\sum_{j \in \mathcal{\hat{S}}_0} a_j P_j)} \cdot e^{it(\sum_{j \in \mathcal{\hat{S}}_1} a_j P_j)} \, ,
\end{align}
as $V_0$ anti-commute with $\hat{\mathcal{S}}_1$ and commute with $\hat{\mathcal{S}}_0$ (if there are terms in $\hat{\mathcal{S}}_0$ anti-commute with $V_0$, we could absorb them into $\hat{\mathcal{S}}_1$ and get a new $\hat{\mathcal{S}}_0$ without these terms).

Then for a Pauli operator $V$ generated by set $\mathcal{W}$, which means it is a multiplication of some terms in  $\mathcal{W}$, then we can get 

\begin{align}
    \nonumber V U V &= e^{-it(\sum_{j \in \mathcal{\hat{S}}_{00}} a_j P_j)} \cdot e^{it(\sum_{j \in \mathcal{\hat{S}}_{01}} a_j P_j)} \cdot e^{-it(\sum_{j \in \mathcal{\hat{S}}_{10}} a_j P_j)} \cdot e^{it(\sum_{j \in \mathcal{\hat{S}}_{11}} a_j P_j)} \\
    V V_0 U V_0 V &= e^{-it(\sum_{j \in \mathcal{\hat{S}}_{00}} a_j P_j)} \cdot e^{it(\sum_{j \in \mathcal{\hat{S}}_{01}} a_j P_j)} \cdot e^{it(\sum_{j \in \mathcal{\hat{S}}_{10}} a_j P_j)} \cdot e^{-it(\sum_{j \in \mathcal{\hat{S}}_{11}} a_j P_j)}\, ,
\end{align}
where $\mathcal{\hat{S}}_{00} \in \mathcal{\hat{S}}_{0}$, $\mathcal{\hat{S}}_{10} \in \mathcal{\hat{S}}_{1}$ be the terms commute with $V$, and $\mathcal{\hat{S}}_{01} \in \mathcal{\hat{S}}_{0}$, $\mathcal{\hat{S}}_{11} \in \mathcal{\hat{S}}_{1}$ be the terms anti-commute with $V$. 
Thus we can get:
\begin{align}
    V U V \cdot V V_0 U V_0 V &= e^{-i2t(\sum_{j \in \mathcal{\hat{S}}_{00}} a_j P_j)} \cdot e^{i2t(\sum_{j \in \mathcal{\hat{S}}_{01}} a_j P_j)}\, .
\end{align}

As we have seen in \cref{thm:3} that the circuit construct by set $\mathcal{W}$ will reverse the unitary with Pauli support anti-commute with it, it is then easily to see that 
\begin{align}
    \nonumber V_{L-1} f_{L-1}(U) U V_{L-1} f_{L-1}(U) &= e^{-i2t(\sum_{j \in \mathcal{\hat{S}}_{0}} a_j P_j)} \cdot V_0 U V_0 \\
    \nonumber &= U_0^\dag \cdot U_0^\dag \cdot U_0 \cdot U_1^\dag \\
    &= U^\dag \, .
\end{align}

\end{proof}

Here we give two examples correspond to \cref{thm:4}. 
The first example is the odd-cycle Ising model as discussed in the main text with 
$H=\sum_{i=0}^{6} a_{i} Z_i Z_{i+1} +\sum_{i=1}^{5} b_i X_i$. 
Here we can get $\hat{S}_0=\{Z_6 Z_0\}$, with $V_0 = Y_1Z_2Y_3Z_4Y_5$ and $V_1 = X_0$. Then the circuit for reversing such unitary is $V_1 U V_0 U V_1 U V_0$. 

The second example is a three-qubit Hamiltonian with Pauli support $\mathcal{S} = \{Z_0X_1Z_2, X_0, X_1, X_2, X_0X_1, X_0X_2, X_1X_2, X_0X_1X_2\}$. This Pauli support can be decomposed into $\mathcal{\hat{S}}_0 = \{X_1, X_0X_2, X_0X_1X_2\}$ and $\mathcal{\hat{S}}_1 = \{Z_0X_1Z_2, X_0, X_2, X_0X_1, X_1X_2\}$. Then we can get $V_0 = Z_0X_1Y_2$ and $\mathcal{W} = \{V_1 = Y_1, V_2 = Y_2\}$, and the circuit for reversing such unitary requires querying gate $U$ for seven times as $V_2 U V_0 U V_1 U V_0 U V_2 U V_0 U V_1 U V_0$.

$\qquad$

\section{Robustness analysis for the unitary inverse protocols}~\label{appendix:robustness}
Here we show the details of the experiment results we did to test the robustness of the inverse protocols when the condition doesn’t fully satisfied. 
We consider $N$-qubit unitaries with Pauli support $\mathcal{S}$ as described in the theorems above, along with their corresponding inverse protocols. Denote $\mathcal{S}^\prime$ as the complement of $\mathcal{S}$, representing the set of Pauli terms not present in the ideal Hamiltonian.

We model the noisy Hamiltonian as
\begin{align}
  H = \sum_{P \in \mathcal{S}} \alpha_P P + \delta \sum_{P \in \mathcal{S}^\prime} \beta_P P,
\end{align}
where $\alpha_P$ and $\beta_P$ are the coefficients of the Pauli operators in $\mathcal{S}$ and $\mathcal{S}^\prime$, respectively. The parameter $\delta$ quantifies the relative noise strength:
\begin{align}
  \delta = \frac{\sum_{P\in \mathcal{S}^\prime} \abs{\beta_P} }{\sum_{P\in \mathcal{S}} \abs{\alpha_P} }.
\end{align}

We generate 10000 random unitaries, where the coefficients $\alpha_P$ and $\beta_P$ are sampled independently from a standard normal distribution. For each instance, we compute the fidelity between the Choi state after running the circuit (applied to the noisy unitary) and the Choi state of the ideal inverse. 
For the situations discussed in the main text, the average fidelities of the inverse protocols under various noise strengths are summarized in \cref{tab:robustness}.

\begin{table}[H]
\centering
\begin{tabular}{l|cccc}
\toprule
\diagbox[width=21em]{Protocol ($N$-qubit)}{$\delta$} & $\quad$ 0 $\qquad$ & 0.001 $\quad$ & 0.01 $\quad$ & 0.1 \\
\midrule
1-slot comb (3-qubit, see \cref{sub@fig:commuting_comb_L1})           & $\quad$ 1.0 $\qquad$ & 0.9999922924 $\quad$ & 0.9992235447 $\quad$ & 0.9261527438 \\
3-slot comb a (3-qubit, see \cref{sub@fig:multislot_example})         & $\quad$ 1.0 $\qquad$ & 0.9999998344 $\quad$ & 0.9999697613 $\quad$ & 0.9970135673 \\
3-slot comb b (3-qubit, see \cref{sub@fig:commuting_example})         & $\quad$ 1.0 $\qquad$ & 0.9999998290 $\quad$ & 0.9999706579 $\quad$ & 0.9970073203 \\
7-slot comb (3-qubit, see \cref{sub@fig:commuting_comb_L3})           & $\quad$ 1.0 $\qquad$ & 0.9999994164 $\quad$ & 0.9999280959 $\quad$ & 0.9928984513 \\
15-slot comb (4-qubit, $L=4$)  & $\quad$ 1.0 $\qquad$ & 0.9999993914 $\quad$ & 0.9998773046 $\quad$ & 0.9876844707 \\
\bottomrule
\end{tabular}
\caption{Average fidelity of the inverse protocols under noisy Hamiltonians. Each dataset consists of 10000 unitaries generated as linear combinations of Pauli operators with random coefficients. The 15-slot example considered Pauli support $\mathcal{S} = \{Y_0,Y_1,Y_2,Y_3,Y_0Y_1,Y_0Y_2,Y_0Y_3,Y_1Y_2,Y_1Y_3,Y_2Y_3,Y_0Y_1Y_2,Y_0Y_1Y_3,Y_0Y_2Y_3,Y_1Y_2Y_3,Y_0Y_1Y_2Y_3\}$. Here, $\delta$ is the relative noise strength added to the coefficients.}
\label{tab:robustness}
\end{table}

\section{Proofs of  Theorems for Unitary Complex Conjugate}~\label{appendix:conjugate}
Here we generalize the above theorems to transform unitary into its complex conjugate, where the key point is to notice that $U^* = e^{-it(\sum_{j \in \mathcal{\hat{S}}_0} a_j P_j) + it(\sum_{j \in \mathcal{\hat{S}}_1} a_j P_j)}$, where $\mathcal{\hat{S}}_0$ is the subset of $\mathcal{\hat{S}}$ with Pauli terms contain odd $Y$ terms, and $\mathcal{\hat{S}}_1$ is the subset of $\mathcal{\hat{S}}$ with Pauli terms contain even $Y$ terms. 
(Note: these results could also be easily extended to unitary transpose, as the ideas are the same so we omit the details here.)

\renewcommand\theproposition{5}
\begin{theorem}\label{thm:51}
For an $N$-qubit unitary $U$ with Pauli support $\cS$, it can be transformed to its complex conjugate with a Pauli operator $V$ such that $U^{*} = V U V$, iff for an arbitrary subset \(\hat{\mathcal S}\subseteq \mathcal{S}\), if $\prod_{i\in \hat{\mathcal S}} P_{i} \sim I$, there are even numbers of Pauli operators in \(\hat{\mathcal S}\) which has even $Y$ terms.
\end{theorem}
\renewcommand{\theproposition}{S\arabic{proposition}}
\begin{proof}
Substitute $U$ with $e^{-it\sum a_j P_j}$, we can get $U^* = e^{it\sum a_j P_j^*}$, 
which means that when $P_j$ has even $Y$ terms the coefficient needs to be negative, and when $P_j$ has odd $Y$ terms the coefficient remains the same. 
We denote that Pauli operators have odd $Y$ terms as \(\hat{\mathcal S}_0\) and that those with even $Y$ terms as \(\hat{\mathcal S}_1\).

If a unitary gate $V$ exists such that $VUV^\dag = U^*$, then we get:
\begin{align}
    U^{*} = e^{-it(\sum_{j \in \hat{\mathcal S}_0} a_j P_j - \sum_{j \in \hat{\mathcal S}_1} a_j P_j)} = e^{-it\sum a_j V P_j V^\dag} \, 
\end{align}
which means that $V$ commutes with \(\hat{\mathcal S}_0\) and anti-commutes with \(\hat{\mathcal S}_1\). 
Since for arbitrary coefficients the above equality holds, $V$ can be chosen to be a Pauli operator. 

Now if there exist a subset \(\hat{\mathcal S} \in \mathcal{S}\) with $\prod_{j\in \hat{\mathcal S}} P_{j} \sim I$, and there are odd numbers of Pauli operators in \(\hat{\mathcal S}\) which has odd $Y$ terms, we can get that: 
\begin{align}
    \nonumber V \Big( \prod_{j\in \hat{\mathcal S}\cap\hat{\mathcal S}_0} P_{j} \cdot \prod_{j\in \hat{\mathcal S}\cap\hat{\mathcal S}_1} P_{j} \Big) V 
    &= \Big( \prod_{j\in \hat{\mathcal S}\cap\hat{\mathcal S}_0} V P_{j} V \Big) \cdot \Big( \prod_{j\in \hat{\mathcal S}\cap\hat{\mathcal S}_1} V P_{j} V \Big) \\
    &= \ \Big( -1 \Big)^{|\hat{\mathcal S}\cap\hat{\mathcal S}_1|} \Big( \prod_{j\in \hat{\mathcal S}\cap\hat{\mathcal S}_0} P_{j} \cdot \prod_{j\in \hat{\mathcal S}\cap\hat{\mathcal S}_1} P_{j} \Big) \, ,
\end{align}
the equality holds iff $|\hat{\mathcal S}\cap\hat{\mathcal S}_1|$ is even. 
\end{proof}

When such $V$ exist, it could also be sovled with the Gaussian elimination-based method. 
As an example, for qubit unitary gate, \(\mathcal{S}=\{X, Y, Z\}\), the subset \(\hat{\mathcal S}\) satisfying $\prod_{j=1} P_{j \in \hat{\mathcal S}} \sim I$ is \(\mathcal{S}\) itself. As there are two Pauli operators in this set which has even $Y$ terms, we can find that $U^{*} = Y U Y$ holds for arbitrary qubit unitary gates. 

There is one observation we want to highlight: Traditionally, people always assume that implementing the $U^*$ of an unknown unitary is inherently simpler than implementing its inverse. However, by rigorously comparing the Hamiltonian conditions required for 1-query realizing $U^\dag$ and 1-query realizing $U^*$, we find that this is not universally true. 
$\mathcal{S} = \{X_0, Z_0, Y_1, Y_0Y_1\}$ is an example that $U^\dag$ can be implemented with a single query, while $U^*$ cannot. 
Note, this analysis could also be extended to unitary transposition, where $U^T$ can be implemented with a single query, while $U^*$ and $U^T$ cannot (e.g. $\mathcal{S} = \{Y_0, Y_1, Y_0Y_1\}$). 
These findings reveal that the relative difficulty of unitary transformations, inverse, conjugate, and transpose, highly related to the structure of $H$, fundamentally overturning prior intuition about their inherent complexities.

$\qquad$

\cref{thm:commuting_comb,thm:commuting_comb2} could also be generalized as stated in the following, where we present them in a general theorem, which is a bit complicated as the situation is quite different: 

\renewcommand\theproposition{6}
\begin{theorem}\label{thm:6}
For transforming $U$ into its complex conjugate, if its Pauli support $\mathcal{S}$ can be decomposed into two subset $\hat{\mathcal{S}}_0$ and $\hat{\mathcal{S}}_1$ commuting with each other and all elements in $\hat{\mathcal{S}}_0$ pairwise commute, 
and if the unitary with support $\hat{\mathcal{S}}_0$ can be transformed into its complex conjugate with a Pauli circuit without ancilla qubit by query $U$ for $Q$ times, 
and there exist a Pauli operator $V_0$ anti-commute with $\hat{\mathcal{S}}_1$ and commute with $\hat{\mathcal{S}}_0$, 
and a Pauli operator $V_0^{'}$ commute with all odd-$Y$-term Pauli operations and anti-commute with all even-$Y$-term Pauli operations in $\hat{\mathcal{S}}_1$ and anti-commute with all odd-$Y$-term Pauli operations and commute with all even-$Y$-term Pauli operations in $\hat{\mathcal{S}}_0$, 
then unitary with Pauli support $\mathcal{S}$ can be transformed into its complex conjugate without ancilla qubits by querying $U$ for $2Q-1$ times. 
\end{theorem}
\renewcommand{\theproposition}{S\arabic{proposition}}
\begin{proof}
The expression is a bit complicated, that we need to first clarify the conditions for $V_0$ and $V_0^{'}$. 
Since $\hat{\mathcal{S}}_0$ commute with $\hat{\mathcal{S}}_1$, we can decompose $U$ into $U = U_0 \cdot U_1$ with 
\begin{align}
    \nonumber U_0 &= e^{-it(\sum_{j \in \mathcal{\hat{S}}_0} a_j P_j)} \\
    \nonumber U_1 &= e^{-it(\sum_{j \in \mathcal{\hat{S}}_1} a_j P_j)} \, .
\end{align}
Then the conditions for $V_0$ and $V_0^{'}$ is equivalent as 
\begin{align}
    \nonumber V_0 U V_0 &= U_0 \cdot U_1^\dag \\
    V_0^{'} U V_0^{'} &= U_0^T \cdot U_1^* \, .
\end{align}

Now if $U_0$ can be transformed into its complex conjugate with a Pauli circuit, we can write down this process as 
\begin{align}
    U_0^{*} = \prod_{j=1}^{Q} V_j U V_j \, .
\end{align}
Then we can get the circuit for getting $U^{*}$ which query $U$ for $2Q-1$ times as 
\begin{align}
    \nonumber U^{*} &= V_0^{'} U V_0^{'} \cdot \prod_{j=1}^{Q} (V_j U V_j) \cdot (V_0 V_j U V_j V_0) \\
    \nonumber &= (U_0^T \cdot U_1^*) \cdot U_0^{*} \cdot U_0^{*} \\
    &= U_0^{*} \cdot U_1^* \, ,
\end{align}
where $\prod_{j=1}^{Q} (V_j U V_j) \cdot (V_0 V_j U V_j V_0) = U_0^{*} \cdot U_0^{*}$ is based on the property that when matrix $A$ and $B$ commute, $e^{A+B} \cdot e^{A-B} = e^{2A}$.


\end{proof}

To help understand, here we give an example:
Given a Pauli support \(\mathcal{S}= \{Y_0, Y_1, Y_2, Y_0Y_1, Y_0Y_2, Y_1Y_2, Y_0Y_1Y_2\}\), 
we can decompose it into $\hat{\mathcal{S}}_0 = \{Y_0, Y_1, Y_0Y_1\}$ and $\hat{\mathcal{S}}_1 = \{Y_2, Y_0Y_2, Y_1Y_2, Y_0Y_1Y_2\}$, 
then we can get $V_0 = X_2$, $V_0^{'} = X_0X_1$, and the circuit for getting $U_0^{*}$ as $U_0^{*} = U X_0 U X_0X_1 UX_1$. 
Based on \cref{thm:6}, we can get the final circuit as $U^* = X_0X_1 U X_0X_1 U X_2 U X_0X_2 U X_2 U X_0X_1X_2 U X_2 U X_1X_2$.


$\qquad$

\section{Transform to unitary complex conjugate with ancilla systems.}~\label{appendix:PQComb_conjugate}

Here we report that through numerical optimization we find some two-qubit examples that the Pauli support doesn't satisfy the conditions in \cref{thm:6} and cannot be transformed into unitary complex conjugate with Pauli circuit without ancilla systems, but can be transformed into unitary complex conjugate with 2 or 3 ancilla qubits. 
These examples show that the required ancilla-qubit number may be reduced,
as previously $d$-dimensional unitary complex conjugate is realized by parallely using $d-1$ gates~\cite{Miyazaki2019}, and two-qubit system then requires $4$ qubits as the ancilla system.

Here we first show the setting of the numerical experiment we did with the PQComb framework~\cite{Mo2025}, which trains a parametrized circuit to get the protocol for unitary complex conjugate. 
In our experiments, we set the training and test sets consist of unitaries generated with the same fixed Pauli support $\mathcal{S}$, but with different randomly chosen coefficients.
For the training set, the coefficients of the Pauli terms in $\mathcal{S}$ are sampled independently from the standard normal distribution $\mathcal{N}(0,1)$. Empirically, we observed that using the normal distribution leads to faster convergence and more effective minimization of the loss function, compared to sampling from a uniform distribution over a fixed interval. In our experiments, protocols trained with normally distributed coefficients are more likely to achieve successful results.

For the test set, we generate unitaries with coefficients sampled independently from a uniform distribution centered at zero. This allows us to evaluate the generalization performance of the trained protocol across a wide range of possible unitaries with the same support. 
This approach ensures that the protocol is robust and performs well not only on the training distribution but also on unseen unitaries with different coefficient distributions.

Through the numerical optimization, we found that $\mathcal{S} = \{ X_1, X_0, Z_0, Y_0, Z_0 X_1, X_0 X_1 \}$ and $\mathcal{S} = \{ Z_0 Z_1, X_0 Z_1, Y_1, Y_0 Y_1, Z_0 X_1, X_0 X_1 \}$ are the examples that their unitary complex conjugate can be realized with 2 ancilla qubits and using gate $U$ sequentially for 3 times. For $\mathcal{S} = \{ Z_0 Y_1, X_0 Z_1, X_0, X_0 Y_1, Z_0, Y_1 \}$ and $\mathcal{S} = \{ Z_0 Y_1, X_1, X_0, Y_0 Z_1, Z_0, Z_0 X_1 \}$, we found that their unitary complex conjugate can be realized with 3 ancilla qubits and using gate $U$ sequentially for 3 times. 
These examples may worth further analysis to get the general conditions and the protocol for realizing their unitary complex conjugate, as reducing the ancilla qubit number has practical value for near-term devices. 
The numerical training process and the final circuit for realizing unitary complex conjugate are given in~\cite{EfficientUnitaryInversion2025}.

$\qquad$

\section{Get linear decomposition for Unitary Transpose}~\label{appendix:transpose}
In this section, we discuss the conditions for linearly decomposing unitary transpose to realize modified Quantum Unitary Reversal Algorithm(QURA), and provide a parameterized optimization method to get the linear decomposition. 

Recall that for arbitrary qubit unitary, its transpose can be linearly decomposed as $IUI+XUX-YUY+ZUZ=2U^T$, thus QURA could use it to reverse $U$ by running the Amplitude Amplification (AA) process of 3 rounds(include the initialization), where the LCU circuit used with this decomposition is shown in \cref{fig:linear_decompose_qubit}. 
Specific Hamiltonians permit dramatic simplifications, e.g. $XUX=U^T$ holds for $H=aX+bY$, it reduces the round of AA to just 1. 

\begin{center}
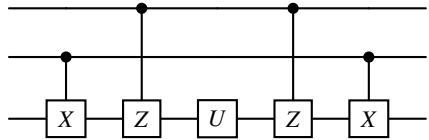
\begin{figure}[htbp]
    \centering
    \begin{quantikz}
        & & \ctrl{2} & & \ctrl{2} & & \\
        & \ctrl{1} & & & & \ctrl{1} & \\
        & \gate{X} & \gate{Z} & \gate{U} & \gate{Z} & \gate{X} &
    \end{quantikz}
    \caption{\centering The LCU circuit used in qubit unitary inverse.}
    \label{fig:linear_decompose_qubit}
\end{figure}
\end{center}

For given $N$-qubit Pauli support $\mathcal{S}$, now the task is to find a set of integer parameters such that 
\begin{align}
    b \cdot U^T = \sum_{i=1}^{4^N} a_i \sigma_i U \sigma_i \, ,
\end{align}
where $\sigma_i$ are the $N$-qubit Pauli operators. 
To make the modified QURA more efficiently, we need to rotate with an angle as large as possible in each round of AA, and the amount of the non-zero $a_i$ be smaller to reduce the ancilla qubit number. 
But there is one more condition which makes it difficult to find the required decomposition with the largest rotating angle and the minimum number of non-zero integers by using classical optimization methods. 
This condition corresponds to the duality based rotating method in QURA, which could be stated as following: 
for arbitary $N$-qubit Pauli operator $\sigma_j$, $b \cdot \sigma_j U^T \sigma_j = \sum_{i=1}^{4^N} a_i \sigma_j\sigma_i U \sigma_i\sigma_j = \sum_{i=k}^{4^N} a_i{'} \sigma_k U \sigma_k$, then $\sum_{i=1}^{4^N} a_i \cdot a_i^{'} = 0$.

To deal with this, we provide a parameterized optimization method based on the PQComb framework~\cite{Mo2025}. 
The idea is that the largest rotating angle in AA corresponds to the highest probability of transforming $U$ into its transpose with a 1-slot quantum comb. 
It inspired to construct a parameterized 1-slot quantum comb with LCU structure. 
After the optimization and get the largest probability, the circuit directly gives us the linear decomposition of $U$ with the given Pauli support.

As an example, we show how to use it for 3-qubit unitary transpose. 
This method could be easily generalized to an arbitrary qubit number. 
Fig.~\ref{fig:linear_decompose} shows the circuit for 3-qubit unitaries, where the variables are $\vec{\theta}_1$ and $\vec{\theta}_2$ in the circuit. 
The lower 3 qubits are the target systems, which after post-selection on the ancilla systems, we wish will perform $U^T$ on the target qubits. 
The upper 6 qubits are ancilla qubits, which are initialized in zero states and finally measured in the computational basis, with $p_{0...0}$ being the success probability. 
The unitary transpose can be realized more efficiently when $p_{0...0}$ is larger. 

In the optimization, we will randomly generated $J$ different unitaries with different coefficients. 
For each generated $U_j$, we calculated the fidelity between the Choi operator of the wanted output $U_j^{T}$, and the real operation $C(U_j)$, denoted as $F(U_j)$, and the probability of getting `000000' as $p(U_j)$. 
The loss function we used here is $l = \alpha\frac{1}{J}\sum_{j=0}^{J-1}(1-F(U_j)) + (1-\alpha)\frac{1}{J}\sum_{j=0}^{J-1}p(U_j)$, where $\alpha$ is a penalty term (we choose $\alpha=0.95$ in running optimization). 
By minimizing this loss function, we get the optimal probability of realizing the unitary transpose, with average fidelity close to 1. 
This circuit then gives directly the linear decomposition of realizing the unitary transpose.

\begin{center}
\begin{figure}[htbp]
    \centering
    \resizebox{0.85\textwidth}{!}{
    \begin{quantikz}
        \lstick[6]{$\ket{0}^{\otimes 6}$} & \gate[6]{V(\theta_1)} &  & & & & & \ctrl{8} & & \ctrl{8} & & & & & & \gate[6]{V(\theta_2)} & & \metercw[6]{000000} \\
        \lstick{} & & & & & & \ctrl{7} & & & & \ctrl{7} & & & & & & &  \\
        \lstick{} & & & & & \ctrl{5} & & & & & & \ctrl{5} & & & & & & \rstick{} \\
        \lstick{} & & & & \ctrl{4} & & & & & & & & \ctrl{4} & & & & & \rstick{} \\
        \lstick{} & & & \ctrl{2} & & & & & & & & & & \ctrl{2} & & & & \rstick{} \\
        \lstick{} & & \ctrl{1} & & & & & & & & & & & & \ctrl{1} & & & \rstick{} \\
        \lstick[3]{\ket{\psi}} & & \gate{X} & \gate[1]{Z} & & & & & \gate[3]{U} & & & & & \gate[1]{Z} & \gate{X} & & & \rstick[3]{$U^T \ket{\psi}$} \\
        \lstick{} & & & & \gate{X} & \gate[1]{Z} & & & & & & \gate[1]{Z} & \gate{X} & & & & & \rstick{} \\
        \lstick{} & & & & & & \gate{X} & \gate[1]{Z} & & \gate[1]{Z} & \gate{X} & & & & & & & \rstick{}
    \end{quantikz}
    }
    \caption{\centering The circuit for getting the linear decomposition of 3-qubit unitary transpose.}
    \label{fig:linear_decompose}
\end{figure}
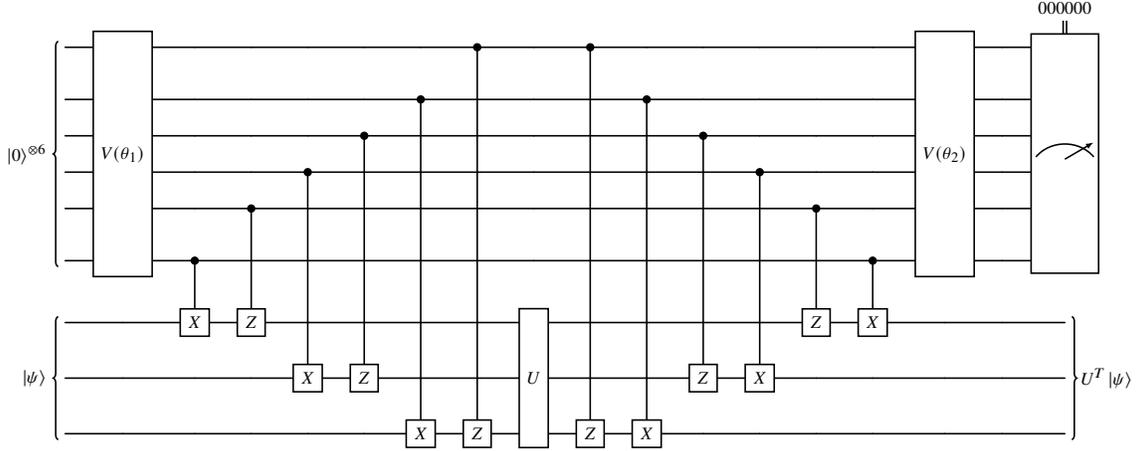
\end{center}

\end{document}